\documentclass[twocolumn,prl,nofootinbib]{revtex4}
\usepackage{amsmath}
\usepackage{graphicx}
\usepackage{dcolumn}
\usepackage{bm}
\usepackage{amssymb}
\usepackage{latexsym}

\def\be{\begin{equation}}
\def\ee{\end{equation}}
\def\bea{\begin{eqnarray}}
\def\eea{\end{eqnarray}}

\begin{document}

\title{Thermal fluctuations of bouncing cosmology revisited}

\author{Taotao Qiu\footnote{qiutt@ihep.ac.cn}}

\affiliation{Physics Department, Chung-Yuan Christian University,
Chung-li, Taiwan 320}

\begin{abstract}
In this note, we revisit the thermal fluctuations generated during
bouncing cosmology, taking Unruh effect into account. We find that
due to the additional effect on temperature, the dependence of power
spectrum on $k$ will get corrected with an indication of blue tilt
at large $k$ region, which is in consistent with the case of vacuum
initial conditions.
\end{abstract}

\maketitle

\section{introduction}
The well-known observational data of Cosmic Microwave Background
anisotropy as well as Large Scale Structure \cite{Komatsu:2010fb}
shows that the structure of our universe, such as galaxies,
temperature fluctuations and so on, comes from the primordial
perturbation at early times with a nearly scale invariant power
spectrum, as successfully predicted by Standard Big Bang Theory
\cite{Mukhanov:1981xt,Guth:1982ec,Hawking:1982cz,Starobinsky:1982ee,Bardeen:1983qw}.
In common sense, this perturbation was initiated from vacuum quantum
fluctuations, which was then stretched out of the horizon. It was
noticed, however, that the vacuum initial condition is by no means
the only way to generate cosmic perturbations and form structure.
Thermal fluctuation, of which fluctuations comes from a thermal
ensemble, has been argued to be the alternative origin to source the
cosmic structure \cite{Magueijo:2002pg}.

Unfortunately, as was concluded in \cite{Magueijo:2002pg} and
summarized in the following paper \cite{Magueijo:2007wf}, that the
scale invariant power spectrum can not be generated in normal
inflationary stage. Due to this reason, people begin to find new
physics in order to save the scale invariance, Among which are
string gas cosmology \cite{Nayeri:2005ck,Brandenberger:2006vv},
non-commutative inflation scenario \cite{Koh:2007rx}, holographic
cosmology \cite{Magueijo:2006fu}, loop quantum cosmology
\cite{Singh:2005km} and so on. See also
\cite{Piao:2007ci,Piao:2007eq,Magueijo:2007na,Li:2010vh}.

Recently, people paid more and more attention to the scenario of
non-singular bouncing cosmology, which has the advantage of getting
rid of the initial singularity which plagued the Standard Big Bang
or inflation theory of cosmology
(\cite{Brandenberger:1988aj,Aref'eva:2007uk,Brustein:1997cv,
Cartier:1999vk,Tsujikawa:2002qc,Biswas:2005qr,Biswas:2006bs,Cai:2009in,
Creminelli:2007aq,Buchbinder:2007ad,Cai:2007qw,Cai:2007zv,Cai:2008qb,
Cai:2008qw}, also see \cite{Novello:2008ra} for a review and the
references therein). Starting from a large and cold initial state, a
bounce scenario can also lead to scale invariant power spectrum of
perturbation, It was then argued in \cite{Cai:2009rd} that thermal
fluctuations is not only possible to occur, but also able to obtain
scale invariant power spectrum as the case of vacuum initial
conditions. The fluctuations can either be ensemble of particle gas,
or holographic radiation, or even string gas. Imprints of
non-Gaussianities could also been obtained due to thermal
fluctuations.

In this paper, we revisit the thermal fluctuations generated during
bouncing cosmologies, taking for example that the fluctuations as
normal particle radiation of which the equation of state
$w_r=\frac{1}{3}$. When the fluctuation moves with varying velocity,
i.e. with an acceleration (which is a very common case), then an
Unruh effect should be taken into account. It will cause an
additional ``Unruh" temperature which may give a correction to the
original temperature $T_r$. We have calculated that it will scale as
$T_r$ times some power law of $k$. For the original case which can
get scale invariant power spectrum, the power law index is positive
and will give a enhancement at large $k$ region (small scales). It
will cause an indication of blue tilt in the power spectrum, which
is in consistent with what happens in the case of vacuum initial
conditions \cite{Cai:2008qw}.

The paper will be organized as follows: we first remind the reader
the main steps in calculating thermal fluctuations in bounce
cosmology in the next section, and then give our results on
correction to the spectrum index by taking the Unruh effect into
account. The last paragraph is the summary. We leave the basic
conjecture for background evolution and the formulae for
perturbation in the Appendix A.






\section{Thermal Fluctuations in bouncing cosmology}

In this section, we review the thermal fluctuations in bouncing
cosmology, as is already mentioned in \cite{Cai:2009rd,Cai:2008qw}.
In order to obtain the spectrum at Hubble crossing at the expanding
phase, we need to calculate it first in the contracting phase, then
transform it into expanding phase using matching conditions at the
bouncing point. We put the general argument of the metric
perturbation evolution and matching conditions to Appendix A. First
of all, from a perturbed Friedmann-Robertson-Walker (FRW) metric,
\be
ds^2=a(\eta)^2[(1+2\Phi)d\eta^2-(1-2\Phi)d\overrightarrow{x}^2]~,\ee
we write down the first perturbed Einstein Equation as: \be -3{\cal
H}(\Phi'+{\cal H}\Phi)+\nabla^2\Phi=4\pi Ga^2\delta\rho~.\ee In the
above two formulae, $a(\eta)$ is the scale factor in terms of the
conformal time $\eta\equiv\int\frac{dt}{a(t)}$, $\Phi$ is the
Newtonian potential, ${\cal H}\equiv\frac{da}{ad\eta}$ is the Hubble
parameter in the conformal time, $\delta\rho$ is the perturbation of
energy density, and prime denotes for derivative with respect to
$\eta$. Assumption is taken that we are in conformal Newtonian gauge
and there is no anisotropic stress \cite{Mukhanov:1990me}. At Hubble
radius crossing all the three terms on the left-hand side of the
above equation are of the same order of magnitude, so modulo a
constant of the order one, we can write down the Fourier space
correlation function of $\Phi$ as: \be\label{Phi2}
|\Phi(k)|^2=16\pi^2G^2k^{-4}a^4<\delta\rho(k)^2>~.\ee

In the system of thermal equilibrium, the energy density
fluctuations are determined by thermodynamics. The correlation
function of $\delta\rho$ in momentum space are correlated to that in
position space by the relation $<\delta\rho(k)^2>\sim
k^{-3}<\delta\rho(x)^2>$, where the latter can be represented as:
\be <\delta\rho(x)^2>=C_V(R)\frac{T^2}{R^6}~,\ee where
$C_V(R)\equiv\frac{\partial<E>}{\partial T}$ is the heat capacity in
a sphere of radius $R$.

We assume that the fluctuations consist of an ensemble of hot
particle gas with constant equation of state $w_r$. Then from the
continuation function and the second thermodynamics law, one can
derive the energy density and temperature which
scale as: \bea \rho_r&\sim&a^{-3(1+w_r)}~,\\
T_r&\sim&a^{-3w_r}~,\eea respectively. This gives \be \rho_r\sim
T_r^{1+\frac{1}{w_r}}~,\ee and thus \bea\label{deltarho2}
<\delta\rho(x)^2>&=&\frac{T_r^2}{R^3}\frac{\partial\rho_r}{\partial
T_r}~\nonumber\\ &\sim&R^{-3}T_r^{2+\frac{1}{w_r}}~\nonumber\\
&\sim&t^{-3(1+p)-6pw_r}~,\eea where in the last formulation we have
used $R=\frac{1}{H}$. Combining Eqs. (\ref{Phi2}) and
(\ref{deltarho2}), one can obtain the dependence on $t$ and $k$ of
$|\Phi|^2$ that is: \be
|\Phi|^2\sim\frac{k^{-3}<\delta\rho(x)^2>}{H^4}\sim
k^{-3}t^{1-3p-6pw_r}~,\ee and \be\label{Phi} |\Phi|\sim
k^{-\frac{3}{2}}t^{\frac{1-3p-6pw_r}{2}}~.\ee

At the Hubble crossing, where $k_\ast=t_\ast^{p-1}$, we have \be
|\Phi|\sim k_\ast^\frac{1-3p(1+2w_r)}{p-1}~,\ee thus the power
spectrum at Hubble crossing time $t_\ast$ will be (See Appendix A
for more details): \be {\cal
P}_\Phi(k_\ast)\equiv\frac{k_\ast^3}{2\pi}|\Phi|^2\sim
k_\ast^{\frac{1-3p(1+2w_r)}{p-1}}~.\ee From above we can see that
the condition for the scale invariance of the spectrum is \be
1-3p-6pw_r=0~,\ee and for the case that the fluctuation behaves like
normal radiation, $w_r=\frac{1}{3}$, it is straightforwardly
obtained that $p=\frac{1}{5}$ and the equation of state of the
background $w=\frac{7}{3}$.


\section{Correction from Unruh effect}
The Unruh effect, which was first described by S. Fulling in 1973,
P. Davies in 1975 and B. Unruh in 1976 \cite{Fulling:1972md},
predicts that an accelerating observer will observe non-zero
temperature where an initial observer would see none. In other
words, from the viewpoint of the observer, particles which is
accelerating will look like a warm gas which is in thermal
equilibrium with non-zero temperature \cite{Bertlmann:2002vi}. As a
consequence, it will cause the black body radiation as Unruh
radiation, which can be viewed as the near-horizon form of Hawking
radiation (see e. g. \cite{Crispino:2007eb} for a comprehensive
review and references therein).

We consider the Unruh effect by taking the velocity part of the
fluctuation into account. The $0-i$ component of the perturbed
Einstein Equations is:

\be (\Phi'+{\cal H}\Phi)_{,i}=4\pi Ga(\rho+P)\delta u_i~,\ee where
$``,i"$ denotes the covariant derivative with respect to the
3-dimensional metric $\gamma_{ij}$, and $\delta u_i$ is the velocity
of perturbation in $i$ direction. Ignoring the anisotropy of the
perturbation and assuming $\delta u_1=\delta u_2=\delta u_3=\delta
u$, and presenting the perturbation in its Fourier form, we have:
\be\label{velocity} |\delta u(k)|=\big|\frac{k(\Phi(k)'+{\cal
H}\Phi(k))}{4\pi Ga(\rho+P)}\big|~.\ee

It is useful to define the comoving curvature perturbation:
\be\label{zeta} \zeta=\Phi+\frac{{\cal H}}{{\cal H}^2-{\cal
H}'}(\Phi'+{\cal H}\Phi)~,\ee which satisfies the equation \be
\zeta''+2\frac{z'}{z}\zeta+c^2_sk^2\zeta=0~,\ee where
$z\equiv\frac{a{\cal H}}{\sqrt{{\cal H}^2-{\cal H}'}}$. Furthermore,
by assuming that in the whole process the non-adiabatic perturbation
can be ignored \footnote{It is a reasonable assumption and can be
realized easily in bounce cosmology if we require the bounce occurs
fast enough. For example, in two fields bouncing model, the short
duration of the bounce relates to small ratio between masses of
light and heavy fields, which can then suppress the isocurvature
perturbation. This can be seen in Ref. \cite{Cai:2008qw}.}, we have:
\be\label{zeta'} \zeta'=-\frac{k^2(\Phi'+{\cal H}\Phi)}{12\pi
Ga^2(\rho+P)}~,\ee which shows that $\zeta$ is conserved at large
scales.

From (\ref{velocity}) and (\ref{zeta'}), and considering the
velocity perturbation in position space, $\delta u(x)\sim
k^{\frac{3}{2}}\delta u(k)$, we have \be |\delta
u(x)|=3ak^{\frac{1}{2}}|\zeta|'~,\ee which relates the dynamics of
the fluctuations to the metric perturbation. In general, the
fluctuation is not moving uniformly and with an acceleration $a_u$.
Thus from above, we can define the Unruh temperature of the
perturbation to be: \be T_u\equiv\frac{a_u}{2\pi}=\frac{\dot{|\delta
u(x)|}}{2\pi}\sim\frac{1}{2\pi}(3a^2k^\frac{1}{2}|\dot\zeta|)\dot{}~,\ee
where a dot means derivative with respect to cosmic time $t$ and we
used $\frac{d}{dt}=\frac{d}{ad\eta}$.

Since all the terms in (\ref{zeta}) are of the same order of
magnitude, then up to a coefficient of order one, we have: \bea
T_u&=&\frac{1}{2\pi}(3a^2k^{\frac{1}{2}}|\dot\Phi|)\dot{}~,\nonumber\\
&\sim&\frac{1}{2\pi}(3t^{2p}k^{\frac{1}{2}}(k^{-\frac{3}{2}}t^{\frac{1-3p-6pw_r}{2}})\dot{})\dot{}~,\eea
where Eq.(\ref{Phi}) is taken into account. Then we have: \be
T_u\sim k^{-1}t^{\frac{p-3}{2}-3pw_r}\sim
k^{-1}t^{\frac{p-3}{2}}T_r~.\ee At the Hubble crossing, we replace
$t$ by $k^{\frac{1}{p-1}}$ to get: \be\label{relation} T_u\sim
k^{\frac{1+p}{2(1-p)}}T_r~.\ee This is the relation between the
temperature of the thermal gas fluctuation and its Unruh correction,
and the relationship depends on the equation of state of the
background\footnote{One may be worry about whether the amplitude of
Unruh temperature can be large enough to really affect the original
temperature $T_r$. We point out that the analysis in this paper is
quite qualitative, and the specific amplitude is determined by the
initial condition of background evolution of the model. Given proper
initial conditions, it is not difficult to let the corrections
comparable to $T_r$. We thank Yi Wang for reminding us this point.}.
For example, for the $w=\frac{7}{3}$ case of which a scale invariant
power spectrum can be obtained, we have
$p=\frac{2}{3(1+w)}=\frac{1}{5}$ and thus $T_u\sim
k^{\frac{3}{4}}T_r$.

This is our main result in this paper. We conclude that the Unruh
Temperature caused by thermal fluctuations scales as the original
temperature times a term that has positive power law of $k$. This
means that in the region of large $k$, the Unruh effect will get
larger and have a magnificant backreaction on the fluctuation.

Next We will roughly estimate how the backreaction would be. From
the above relation (\ref{relation}) between $T_r$ and $T_u$, we can
similarly parameterize $T_u\sim a^{-3w_r'}$, where one can calculate
easily that at the Hubble crossing $w_r'=w_r-\frac{p-1}{4p}$. At the
large $k$ region where $T_u$ dominates, one can get the dependence
of the power spectrum on $k$ as what has be done at the above,
${\cal P}_\Phi(k)\sim k^{\frac{1-3p-6pw_r'}{p-1}}$. For our case
$p=\frac{1}{5}$ and $w_r=\frac{1}{3}$, we can get
$w_r'=\frac{4}{3}$, and thus ${\cal P}_\Phi(k)\sim k^{\frac{3}{2}}$.
From this we can see, at the large $k$ region, the power spectrum
will get an indication of blue tilt, which is in consistent with the
case of vacuum initial conditions.






\section{summary}
In this short note, we reinvestigated the thermal fluctuations
generated during bouncing cosmology, taking into account the
corrections from Unruh effect. Our calculation shows that for the
original case of scale invariant power spectrum, the additional
correction leads to a blue tilt at large $k$ region, which is in
consistent with the case of vacuum initial condition as previously
discussed.

Unruh effect is of great importance in its own right and has been
applied as a tool to investigate other phenomena such as the thermal
emission of particles from black holes
\cite{Hawking:1974rv,Hawking:1974sw} and cosmological horizons
\cite{Gibbons:1977mu}. However, its applications on thermodynamics
of cosmological perturbations is not clear yet. Whether it will give
rise to more corrections or modifications to the early universe is
interesting and worth studying. It can also been applied to the
Holographic Cosmology, cf. the recently released paper by E.
Verlinde \cite{Verlinde:2010hp} who argues that the Einstein Gravity
is emergent from thermodynamics and holographic principle. These
extensions of this project are expected to be done in the future
work.

\section*{Acknowledgments}
The author would thank Yifu Cai, Shi Pi, Yi Wang, Prof. Yunsong Piao
and Prof. Chiang-Mei Chen's group for useful discussion. This
research is supported in parts by the National Science Council of
R.O.C. under Grant No. NSC96-2112-M-033-004-MY3 and No.
NSC97-2811-033-003 and by the National Center for Theoretical
Science.

\appendix
\section{Appendix A}

The appendix aims at clarifying the basic foundations of background
and perturbation evolution for our analysis. In the framework of
Friedmann-Robertson-Walker(FRW) geometry and taking the assumption
of some constant equation of state $w$ for the background, the scale
factor of the universe scales as: \be a\sim t^p~,\ee where \be
p=\frac{3(1+w)}{2}~,\ee and the Hubble parameter becomes: \be
H=\frac{p}{t}~.\ee Thus from the conditions for the comoving wave
mode $k$ at Hubble crossing, we have \be
k_\ast=a(t_\ast)H(t_\ast)\sim t_\ast^{p-1}~.\ee

Then we briefly review the perturbation evolution through the
bounce. From the perturbed Einstein equation, we can get the
equation for $\Phi$ as: \be \Phi''+k^2\Phi+\frac{2{\cal H}^3-4{\cal
H}{\cal H}'+{\cal H}''}{{\cal H}^2-{\cal H}'}\Phi'+\frac{{\cal
H}{\cal H}''-2{\cal H}'^2}{{\cal H}^2-{\cal H}'}\Phi=0~.\ee The
equation has a general solution of $\Phi(k,\eta)$ which can be
parameterized as: \be
\Phi(k,\eta)^{\pm}=D(k)_{\pm}+S(k)_{\pm}\eta^{-2\nu}~,\ee where
$\nu=\frac{5+3w}{2(1+3w)}$ and $\pm$ refers to modes in the
expanding (contracting) phase. Applying the Hwang-Vishniac
\cite{Hwang:1991an} (Dueruelle-Mukhanov \cite{Deruelle:1995kd})
conditions which match modes before and after the bounce, we can get
the dominant mode for the expanding phase turns out to be
\cite{Lyth:2001pf,Lyth:2001nv,Brandenberger:2001bs,Allen:2004vz}:
\be D_+={\cal O}(1)D_-+{\cal O}(1)(\frac{k}{k_\ast})^2S_-~.\ee

The spectrum in the expanding phase can only be given by the $D_-$
mode \cite{Cai:2009rd}, which indicates that \be {\cal
P}_\Phi(k)\sim{\cal P}_{D_-}(k)~.\ee

\end{document}